\documentclass[conference]{IEEEtran}
\IEEEoverridecommandlockouts
\usepackage{cite}
\usepackage{amsmath,amssymb,amsfonts}
\usepackage{algorithmic}
\usepackage{graphicx}
\usepackage{textcomp}
\usepackage{xcolor}
\usepackage{subfigure}
\usepackage{parskip}
\def\BibTeX{{\rm B\kern-.05em{\sc i\kern-.025em b}\kern-.08em
    T\kern-.1667em\lower.7ex\hbox{E}\kern-.125emX}}
\newlength{\bibitemsep}
\newlength{\bibparskip}
\let\oldthebibliography\thebibliography
\renewcommand\thebibliography[1]{%
  \oldthebibliography{#1}%
  \setlength{\parskip}{\bibitemsep}%
  \setlength{\itemsep}{\bibparskip}%
}

\begin{document}
\title{Advance Detection Of Bull And Bear Phases In Cryptocurrency Markets Through Technical Analysis}

\makeatletter
\newcommand{\linebreakand}{%
  \end{@IEEEauthorhalign}
  \hfill\mbox{}\par
  \mbox{}\hfill\begin{@IEEEauthorhalign}
}
\makeatother

\author{\IEEEauthorblockN{Rahul Arulkumaran}
\IEEEauthorblockA{\textit{University At Buffalo} \\
Buffalo, New York}
\and \and
\IEEEauthorblockN{Suyash Kumar}
\IEEEauthorblockA{\textit{University At Buffalo} \\
Buffalo, New York}
\and \and
\IEEEauthorblockN{Shikha Tomar}
\IEEEauthorblockA{\textit{University At Buffalo} \\
Buffalo, New York}
\linebreakand
\IEEEauthorblockN{Manideep Gongalla}
\IEEEauthorblockA{\textit{University At Buffalo} \\
Buffalo, New York}

\and
\IEEEauthorblockN{Harshitha}
\IEEEauthorblockA{\textit{University At Buffalo} \\
Buffalo, New York}

}

\parskip=6pt
\title{Advance Detection Of Bull And Bear Phases In Cryptocurrency Markets}
\maketitle
\begin{abstract}
Cryptocurrencies are highly volatile financial instruments with more and more new retail investors joining the scene with each passing day. Bitcoin has always proved to determine in which way the rest of the cryptocurrency market is headed towards. As of today, Bitcoin has a market dominance of close to 50\%.

Bull and bear phases in cryptocurrencies are determined based on Bitcoin’s performance over the 50-Day and 200-Day Moving Averages.

The aim of this paper is to foretell the performance of bitcoin in the near future by employing predictive algorithms. This predicted data will then be used to calculate the 50-Day and 200-Day Moving Averages and subsequently plotted to establish the potential bull and bear phases.
\end{abstract}

\section{Introduction}
A well known fact about the Cryptocurrency markets is that the markets are extremely volatile. Bitcoin was the first cryptocurrency that was developed by Satoshi Nakamota. Satoshi Nakamota is the pseudo name of the person who developed cryptocurrencies. However, until today nobody knows who is Satoshi Nakamoto.

Although the identity of the person who created Bitcoin is unknown, the community did not let that bother them. The aim of Bitcoin was to be truly decentralised as Satoshi Nakamoto mentioned in the whitepaper for Bitcoin and the community embraced this change. Bitcoin was created in 2009 and it has been 11 year since then, but it was only in 2013 that Bitcoin started gaining traction. Retail investors were initially reluctant to invest in Bitcoin but that changed in 2013. Since then, the cryptocurrency markets have seen the advent of many altcoins like Ethereum, Cardano, Ripple, Litecoin, Bitcoin Cash, and many more.

Since then, the price of Bitcoin has skyrocketed and as of writing this paper, Bitcoin's all-time high has been at \$69,000. Although Bitcoin has seen exponential gains in the last couple of years, Bitcoin has always had bullish and bearish cycles. Usually, each cycle lasts for about 4 years and this has allowed most retail and institutional investors to perform robust technical analysis to understand potential price targets in each cycle.

Although Bitcoin has these 4 year cycles, the price of Bitcoin does not always keep increasing or decreasing in bull and bear cycles respectively. There are corrections in bull cycles which result in major pull back in the price and at the same time there are major price pumps during bearish cycles.

With more and more institutional investors entering the cryptocurrency markets, it becomes increasingly difficult to predict possible price targets. Institutional investors usually have the financial power and political support to manipulate markets for their gains. In recent times, Bitcoin too has seen major market manipulations. However, an interesting thing to note is that, even during such manipulations, technical analysis targets are usually met.

Technical Analysis has time and again proven to be a useful method to predict future market scenarios and understand current market trends. Moving Averages are on of the most key technical indicators in the cryptocurrency markets. Moving Averages take into consideration past price context to give traders and investors a sense of expected price at a particular instant.

Moving averages can be calculated over various time instances. The most commonly used Moving Averages are 50-day and 200-day Moving Averages. The 50-day Moving Average is also known as the "Short Term Moving Average" whereas the 200-day Moving Average is also known as the "Long Term Moving Average"

A general concept of seeing crossovers of these 2 moving average lines gives investors a better understanding of current market scenarios. It is generally believed that if the short-term moving average goes crosses and goes below the long-term moving average, then it is a bearish scenario for the markets. This is also called the "Death Cross".

\begin{figure}[htbp]
\centerline{\frame{\includegraphics[scale=0.4]{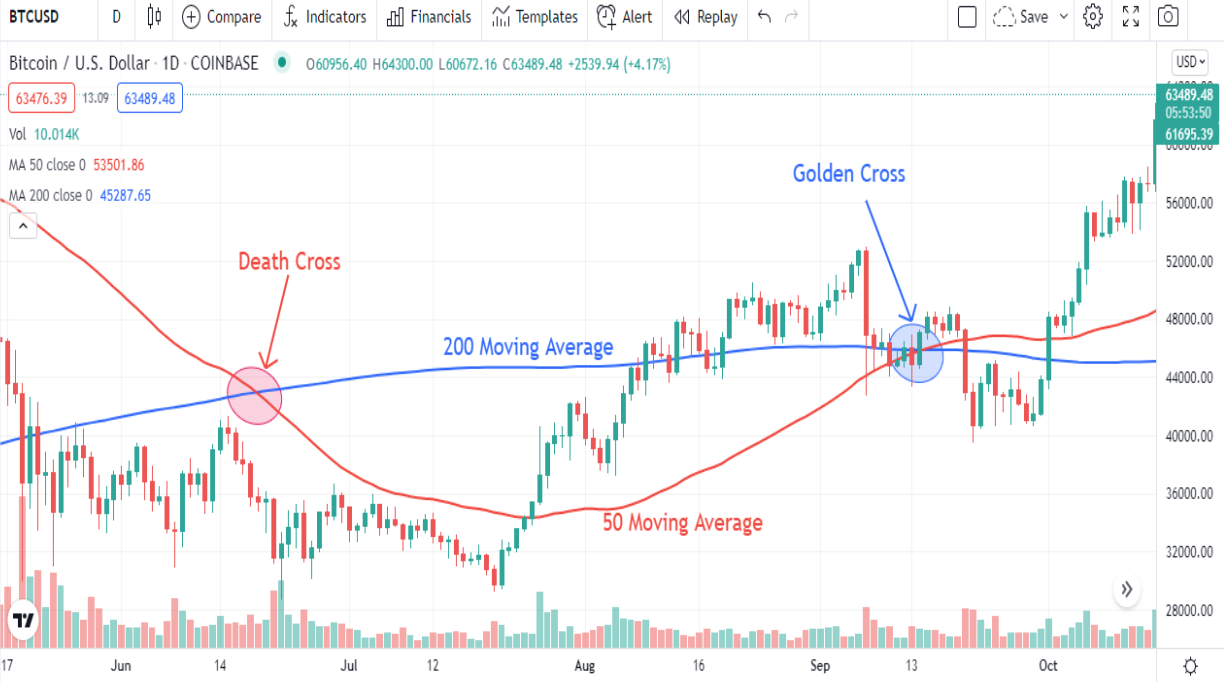}}}
\caption{Graphical Representation Of Golden Cross and Death Cross.}
\label{fig}
\end{figure}

If the long-term moving average goes crosses and goes below the short-term moving average, then it is a bullish scenario for the markets. This bullish scenario is also called the "Golden Cross". 
That being said, Moving average indicators are generally lagging indicators. It takes a while for this indicator to show the true nature of the market. The current market scenario is usually factored into Moving Averages and reflects only a few days later.

In this paper, the authors formulate a mechanism to predict in current day the future prices of the market and then detect then phase in which the market is headed towards. This advance prediction and detection mechanism will help retail investors understand how the market could move in the future.

\section{Predictive Modelling And Advance Phase Detection }
\subsection{Data Collection}
The data collection part was the most important aspect of the project. Data for Bitcoin was gathered from an open API where the Open, High, Low, Close and Volume data were available for each day from 1st January 2012. These features were sufficient to build a robust dataset.

The dataset did not have all the features required like technical indicators. However, the OHLCV data was sufficient to manually calculate the technical indicators required.

\subsection{Data Generation And Technical Indicators}
After collecting the data, the aim was to generate the technical analysis indicators. There are a plethora of technical indicators available to use. The authors in the paper chose Moving Average, RSI, MACD, Momentum, Bollinger Bands and ROC.

Each indicator has its own purpose. RSI talks about whether a particular asset is either overbought or oversold.

MACD is a trend-following momentum indicator that depicts the relationship between two moving averages of an asset’s price.

Momentum indicator depicts the direction in which the price of an assets seems to be moving in.

ROC is an oscillator that fluctuates above and below the zero line. When the price rises, the ROC moves up and when the price declines, the ROC falls.

The closing prices of 21 days in the future were also added to each and as a result 21 new columns were created for this purpose.

\subsection{Data Pre-processing and Exploratory Data Analysis}
After generating all necessary data points, the data was required to be processed. Each of these technical indicators did not have some values in the beginning. This is because, these technical indicators required a minimum number of past closing prices to give accurate results. Rows that had no values in certain columns in the beginning of the dataset were discarded.

Once the data was processed and had all required features, some basic exploratory data analysis was performed. On performing this, a key aspect to address was the high correlation between the features in the dataset. Although each feature had its own importance, the features were all being derived from the existing columns in the dataset.

However, this was not a cause of concern as the aim of our model was to predict the right trends. Finding accuracy of the our final model would not give us an accurate overview of the model and its performance. Instead, the authors decided to plot and check the predicted moving average graphs and see how closely related they are to the actual moving average graphs during that duration and whether they followed trends or not.

\subsection{Model Formulation And Data Splitting}
The authors of this model developed a key method to predict prices of the future. To predict prices, one would require all features for that particular day in the future. However, in current day there is no way to get future input values.

As a result, the authors added the future 21-day closing prices to each day allowing the model to get future context information in present day so that it becomes possible to accurately predict the bitcoin price upto 21 days in the future. The output variables in this case would be the columns with closing prices of 21 days. The input features consisted of OHLCV points, RSI, MACD, Bollinger Bands and ROC.

\begin{figure}[htbp]
\centerline{\frame{\includegraphics[scale=0.29]{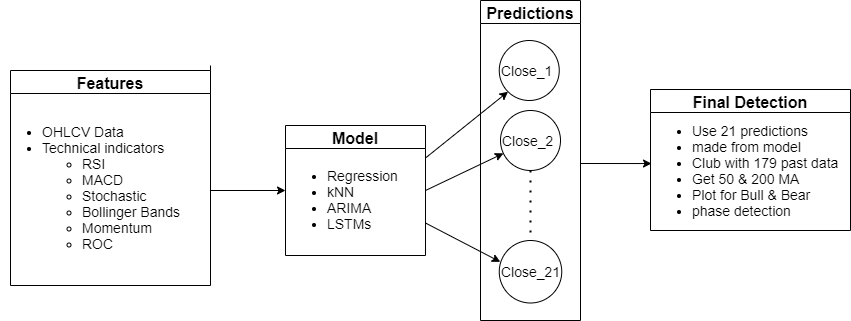}}}
\caption{Project Architecture}
\label{fig}
\end{figure}

For example - Consider taking data from 1st August 2021 to 30th September 2021. First we will add 21 columns, each containing the closing price of the next 21 days and this will be possible to add until 9th September. From 10th September onwards there will not be sufficient data for those 21 columns with closing prices of the future so they will be discarded for training purposes.

This way  a full fledged dataset with all necessary columns filled in each of the rows until 9th September can be built. This should be sufficient for training the model.

For the purpose of comparision the authors built 2 different models - Multiple Linear Regression and LSTM. The data for each of the models was split as follows

\begin{table}[htbp]
\caption{Data Splitting For Training And Testing}
\begin{center}
\begin{tabular}{|c|c|c|}
\hline
\textbf{Model} & \textbf{\textit{Training \%}}& \textbf{\textit{Test \%}} \\
\hline
MLR & 75 & 25   \\
\hline
LSTM & 75 & 25   \\
\hline
\end{tabular}
\label{tab1}
\end{center}
\end{table}

\subsection{Formulation and Implementation of Multiple Linear Regression}
Multiple linear regression (MLR) is used to determine a mathematical relationship among several random variables by checking how they are related to the response variable. After the relationship is deployed between the independent and dependent variables, the same relationship was used to accurately predict the level of effect it had on the outcome variable. MLR does this by trying to fit a linear line that best approximates all the data points.
Below is the equation for the calculation of MLR:

\begin{equation}
y = \beta_0 + \beta_1x_1 +\beta_2x_2 +...+\beta_ix_i +\epsilon
\end{equation}

Here $y$ is the dependent variable, $x$ are the exploratory variable, $\beta_0$ is the y-intercept, $\beta_i$ are the slope coefficient for the exploratory variable and $\epsilon$ is the model error term.
The accuracy of the model is decided by the coefficient of determination R\textsuperscript{2} which varies between 0 and 1, 0 indicating independent variables that have no effect on the dependent variable and 1 indicating dependent variable that can be accurately predicted by the independent variables.

But there can be cases where model is found to be over fitted which can result in high value of R\textsuperscript{2}. One of the reasons this can occur is when high multicollinearity is found between the independent variables.

While building the MLR model for this project, the authors faced two major challenges. First one was the high correlation between the independent features since the technical indicators are derived from the same OHLCV data. The second challenge was regarding the output as there were 22 dependent variables and 7 independent variables, which is not possible in MLR. Thus, the authors used the  MTR(Multiple Target Regression) model wherein multiple target properties can be used to train the model but while predicting the dependent variables using this model resulted in the average of all the target properties rather than the individual values. To circumvent this, the authors went back to MLR and created 22 individual models for each of the 22 individual dependent variables using the same 7 features. 

Before developing the MLR model, dataset was divided into training and testing data, and as this was time series data, training data couldn't be selected randomly, so the latest of the 25\% of data was chosen as testing set while remaining as the training set. The authors trained 22 models on training data for all the 22 closing prices(close, close1..close21) using the 7 features. After the model training was complete, prediction of closing prices was done on testing data and, then, the 50 days and 200 days simple moving averages were calculated.

The authors plotted the actual 50day SMA, 200day SMA and predicted 50day SMA, 200day SMA to check how was the accuracy of the model and if the predicted values were following the same curve as the actual values. This way of checking the accuracy overcame the over fitting and high R\textsuperscript{2} value issue.

\subsection{Formulation and Implementation of LSTM}
The model was formulated and built after splitting the data. Since the data was time-dependent, it was necessary to send context information through to future points. Any pattern does not just appear suddenly. The pattern forms gradually, and as a result, it becomes increasingly essential to pass past context to future inputs. Past context information can be passed through various forms of  Recurrent Neural Networks (RNNs) the most prominent among which being, Long Short Term Memory (LSTMs) networks and Gated Recurrent Units (GRUs). We decided to use LSTMs based on past experiences.

\begin{figure}[htbp]
\centerline{\frame{\includegraphics[scale=0.4]{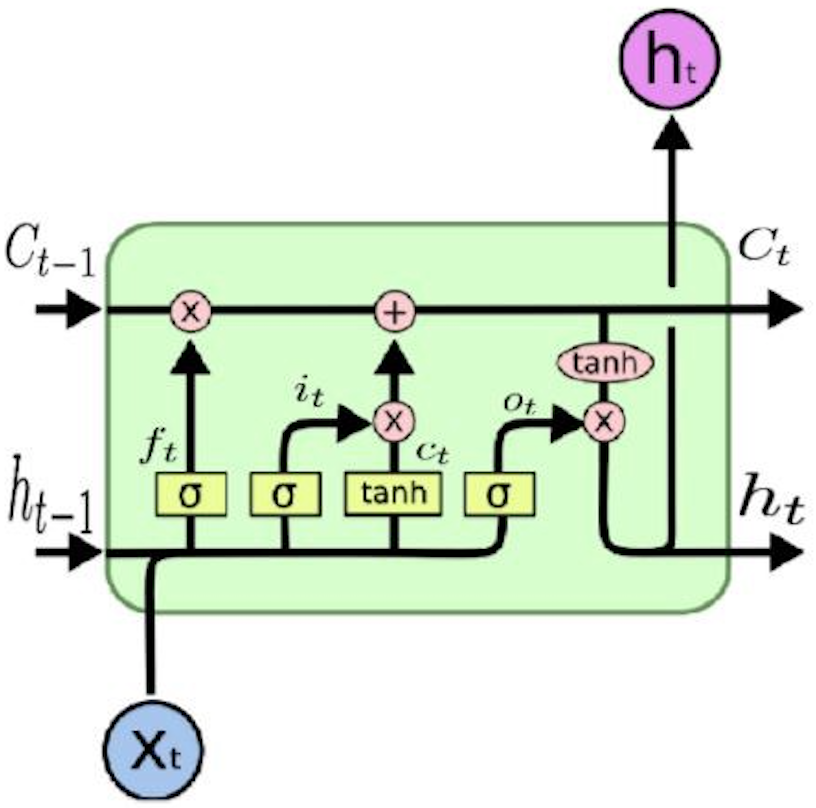}}}
\caption{Predicted 50-day and 200-day Moving Average By LSTM.}
\label{fig}
\end{figure}

The figure above represents one block at instant $t$ and $C_{t-1}$ and $C_{t}$ represent block states at time t-1 and t respectively. $h_{t-1}$ and $h_{t}$ represent the outputs from the LSTM at time t-1 and t respectively and $x_{t}$ represents input vector to the LSTM at time t. 

LSTMs have 3 main parts, namely, Forget Gate, Input Gate, and Output Gate.  All the gates in LSTMs are similar to the gates in digital electronics.
\begin{equation}
f_t = \sigma(W_f.[h_{t-1},x_t] + b_f)
\end{equation}
The Forget Gate enables the LSTM to discard unnecessary information from the past while only retaining necessary past context information. In (2) $W_{f}$ and $b_{f}$ are weights of the Forget Layer and $\sigma$ is the logistic function. The logistic output $f_{t}$ of this layer as depicted in Fig.2 serves as a "Gate" on the $C_{t-1}$ values. When 1, the gate is fully open and when 0, the gate is fully closed. So in the training process, the weights $W_{f}$ and $b_{f}$ adjust themselves to values that allow the right degree of memory from the past to affect the present and future values.
\begin{equation}
i_t = \sigma(W_i.[h_{t-1},x_t] + b_i)
\end{equation}
\begin{equation}
c_t = tanh(W_c.[h_{t-1},x_t] + b_c)
\end{equation}

The input gate controls the extent to which new values are introduced into the LSTM. In (3) $W_{i}$ and $b_{i}$ are weights of the Input Layer and $i_{t}$ its output. Unlike the Forget Gate, the Input Gate is not a standalone gate. It only regulates the fraction of $c_{t}$ allowed to flow into the conveyor belt. The value of $c_{t}$ is calculated as shown in (4) with $W_{c}$ and $b_{c}$ as its corresponding weights. Effectively, the Input Gate relates to two hidden layers and not one.
\begin{equation}
C_t = f_{t}*C_{t-1} + i_{t}*c_{t}
\end{equation}
Equation (5) does not represent another gate. It denotes the amount of information conveyed out from the block at instant $t$ and allowed to impact the future blocks.

\begin{equation}
o_t = \sigma(W_o.[h_{t-1},x_t] + b_o)
\end{equation}
\begin{equation}
h_t = o_{t}*tanh(C_{t})
\end{equation}

The output gate denotes the output of the block. In (6) $W_{o}$ and $b_{o}$ are weights of the Output Layer and $o_{t}$ its output. The output of the block which is the output of the hidden layers ($C_{t}$) passed through the activation functions. But before being passed out as the final output, it is regulated as shown in (7) by $o_{t}$.

The LSTM built for the purpose of achieving the goal mentioned in this paper has two hidden layer, one input layer,  and one output layer. The input layer composed of 100 neurons, while each of the 2 hidden layers composed of 15 and 31 neurons respectively. The activation function used in the hidden layers was “ReLU.” Finally, the output layer just had 22 neurons, which gave a "ReLU" output since the activation of the output layer was "ReLU".

Since "ReLU" is an unbounded function, it aligned with the ideaology mentioned in this paper. Using a sigmoid or a tanh activation function would have been appropriate for a classification model as they are bounded between 0 and 1 and -1 and 1 respectively. "ReLU" does not have an upper bound and aligns perfectly for regression based predictive models. As the aim of the project was to determine a price according to market conditions, the "ReLU" activation function was used.

\begin{table}[htbp]
\caption{Sequential Model Description}
\begin{center}
\begin{tabular}{|c|c|}
\hline
\textbf{Layer} & \textbf{\textit{Output Shape}} \\
\hline
lstm1 (LSTM) & (None, 7)   \\
\hline
dense1 (Dense) & (None, 15)  \\
\hline
dense2 (Dense) & (None, 31)   \\
\hline
dense1 (Dense) & (None, 22)  \\
\hline
\end{tabular}
\label{tab1}
\end{center}
\end{table}

\section{Results and Discussions}

After making the price predictions, the predicted values were then used to compute the Moving Averages. The graph below depicts the actual MA of Bitcoin over the 50-day and 200-day time frames.

\begin{figure}[htbp]
\centerline{\frame{\includegraphics[scale=0.4]{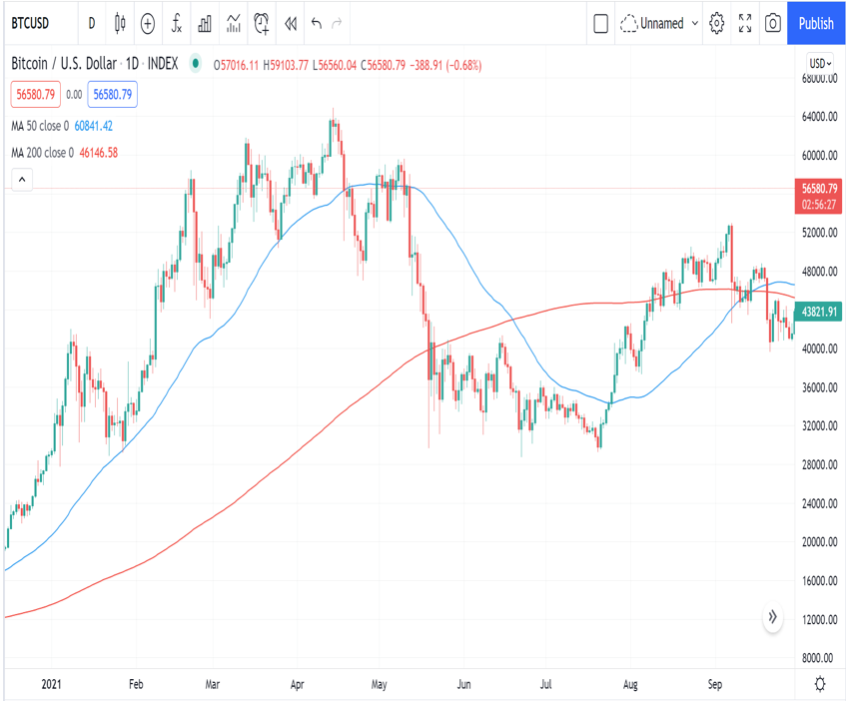}}}
\caption{Actual 50-day and 200-day Moving Average.}
\label{fig}
\end{figure}

The Multiple Linear Regression Model was the first model tested by the models to establish a baseline model. The results produced by the model were not following similar trend as the actual moving averages.

The moving averages computed from the prices predicted by the MLR Model are as depicted below. The MLR model involved a lot more work mainly because 22 different models were created to predict 22 days' closing prices. This would not be effective and efficient if the model would have to extended to predict market conditions for longer a time frame into the future.

The MLR model seemed to be taking into consideration a lot of information from way back in the past rather than giving more emphasis to prices in the recent past. This resulted in prices being slightly more divulged from the actual prices.

The moving averages computed from the predicted prices of the MLR model are as depicted below.
\begin{figure}[htbp]
\centerline{\frame{\includegraphics[scale=0.4]{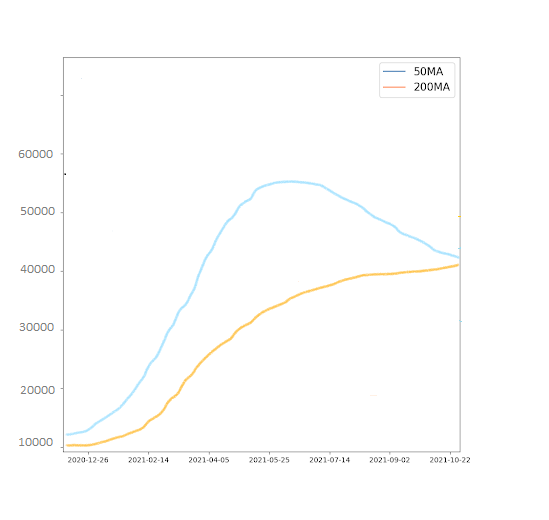}}}
\caption{Predicted 50-day and 200-day Moving Average By MLR.}
\label{fig}
\end{figure}

An interesting thing to note here is that the results shown above were for the LSTM that was trained over 2000 epochs. On training it with 1000 epochs, it was noticed that the results were not as accurate as shown above. Due to the unavailability of time and computations resources, the model could not be trained for beyond 2000 epochs.

The predicted prices using the LSTM were then used to compute the moving averages and the results achieved are as depicted below.

\begin{figure}[htbp]
\centerline{\frame{\includegraphics[scale=0.4]{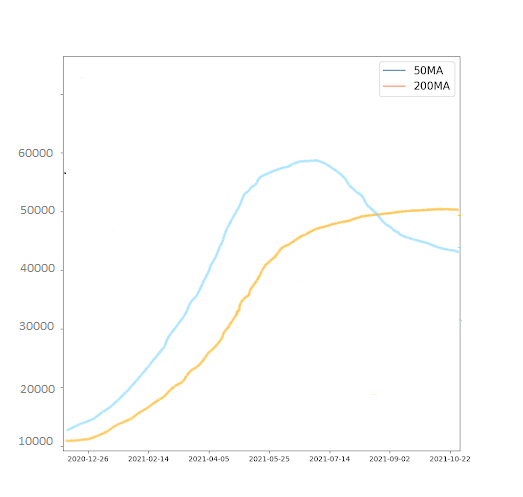}}}
\caption{Predicted 50-day and 200-day Moving Average By LSTM.}
\label{fig}
\end{figure}

However, if the model is trained further the results that the model produces could be more accurate. Finding the point up to which the model can be trained is a key aspect and acts as a hyper parameter. If the model is trained for more epochs than required, the problem of overfitting could arise.

The MLR model was slightly inaccurate as compared to the LSTM model to predict prices. This is most likely because of the fact that the LSTM model was able to establish more closely the combinatorial relationship between various features and mapped it accurately to a price. LSTMs are known to perform well over time-series data mainly because of its architecture. Not all information from the past carries forward to the future and only limited information context from past is sent through. Any unimportant information is dropped by the Forget Gate in the LSTM. However, in MLR the model seems to taking into consideration a lot of information from the past.

This is not necessarily needed as the price of Bitcoin would depend more on traits and characteristics of the market in the recent past rather than the longer past.

The R\textsuperscript{2} score, also known as the coefficient of determination is the proportion of variance in the dependent  variable that  can be predicted from the independent  variables. In the project, the authors chose not to use R\textsuperscript{2} as their accuracy indicator because most of the technical indicators are derived from OHLC price points and, therefore, were highly correlated. Instead, to verify accuracy, the original and predicted Moving Averages graphs were compared.

\section{Conclusion}
Advance detection of market phases in continuous-time financial systems, if attained successfully, will be of great value to investors and traders, and in particular to critical financial processes. Machine Learning techniques, more precisely the “learning-functionality-from-data” methodology of supervised learning, is the most appropriate mechanisms for attaining the same. Previous works have shown that conventional feed-forward Artificial Neural Networks (ANNs) have been able to achieve a fair degree of success in this direction. However, both these techniques work on taking parameter inputs at a single time step of a running process.

Here, LSTMs have been used to extract functionality from time-sequences, with the hypothesis and expectation that more intricate functional relationships between combinatorial patterns can be extracted, leading to achieving greater success in the above objective. This hypothesis has been upheld, as shown in the previous sections.

\vspace{12pt}
\end{document}